\documentclass[aps,prl,showpacs,twocolumn,superscriptaddress]{revtex4}

\usepackage{graphicx}
\usepackage{mathptmx}

\begin{document}

\title{Colossal magnetostriction and negative thermal expansion in
the frustrated antiferromagnet ZnCr$_2$Se$_4$}

\author{J. Hemberger}
\author{H.-A. Krug von Nidda}

\affiliation{Experimental Physics V, Center for Electronic
Correlations and Magnetism, University of Augsburg, D-86159
Augsburg, Germany}

\author{V. Tsurkan}

\thanks{Corresponding author}

\affiliation{Experimental Physics V, Center for Electronic
Correlations and Magnetism, University of Augsburg, D-86159
Augsburg, Germany}

\affiliation{Institute of Applied Physics, Academy of Sciences of
Moldova, MD-2028 Chi\c{s}in\v{a}u, R. Moldova}

\author{A. Loidl}

\affiliation{Experimental Physics V, Center for Electronic
Correlations and Magnetism, University of Augsburg, D-86159
Augsburg, Germany}

\begin{abstract}
A detailed investigation of ZnCr$_2$Se$_4$ is presented which is
dominated by strong ferromagnetic exchange but orders
antiferromagnetically at $T_\mathrm{N}$ = 21~K.  Specific heat $C$
and thermal expansion $\Delta L/L$ exhibit sharp first-order
anomalies at the antiferromagnetic transition. $T_\mathrm{N}$ is
strongly reduced and shifted to lower temperatures by external
magnetic fields and finally is fully suppressed suggesting a field
induced quantum critical behavior close to 60~kOe. $\Delta L/L(T)$
is unusually large and exhibits negative thermal expansion below
75~K down to $T_\mathrm{N}$ indicating strong frustration of the
lattice. Magnetostriction $\Delta L/L(H)$ reveals colossal values
(0.5x10$^{-3}$) comparable to giant magnetostriction materials.
Electron-spin resonance, however, shows negligible spin-orbital
coupling excluding orbitally induced Jahn-Teller distortions. The
obtained results point to a spin-driven origin of the structural
instability at $T_\mathrm{N}$ explained in terms of competing
ferromagnetic and antiferromagnetic exchange interactions yielding
strong bond frustration.
\end{abstract}


\maketitle

Recently magnetic oxides and chalcogenides crystallizing in the
spinel structure have attracted considerable attention. Within the
last few years exotic phenomena and fascinating ground states have
been observed in this class of materials: heavy-fermion behavior
\cite{kondo:97}, complex spin order and spin dimerization
\cite{lee:02,radaelli:02,schmidt:04}, spin-orbital liquid
\cite{fritsch:04} and orbital glass \cite{fichtl:05}, as well as
coexistence of ferromagnetism and ferroelectricity
\cite{hemberger:05,yamasaki:06}. They are attributed to the
cooperativity and competition between charge, spin and orbital
degrees of freedom, all of which are strongly coupled to the
lattice. In addition, topological frustration due to the
tetrahedral arrangement of the magnetic cations and bond
frustration due to competing ferromagnetic (FM) and
antiferromagnetic (AFM) exchange interactions hamper any simple
spin and orbital arrangement in the ground state. Spin-lattice
coupling plays an important role in releasing frustration by
structural transformation. For example, in antiferromagnetic
chromium-oxide spinels frustration in the spin sector due to
geometrical constraints is released by a Peierls-like structural
transition, which has been explained in terms of a so called spin
Jahn-Teller effect \cite{lee:00,tchernyshyov:02,yamashita:00}.
Another source of a structural instability was recently identified
in AFM ZnCr$_2$S$_4$ and ascribed to competing FM and AFM exchange
of nearly equal strength yielding strong bond frustration
\cite{hemberger:06}.

Here we present the results of a study of the spin-lattice
correlations of another zinc chromium spinel, ZnCr$_2$Se$_4$,
which shows an antiferromagnetic ground state despite the presence
of strong FM exchange. Previous neutron-diffraction investigations
revealed a complex antiferromagnetic order  at temperatures below
20 K contrasting with the dominating FM interactions evidenced by
a large positive Curie-Weiss temperature of 115~K
\cite{lotgering:65, plumier:66}. This testifies to the importance
of the next-nearest neighbor $(nnn)$ exchange besides the nearest
neighbor $(nn)$ exchange interactions \cite{lotgering:65}. The
spin structure is incommensurate having a ferromagnetic
arrangement in the (001) planes with a turning angle of 42$^\circ$
between the spins in the adjacent (001) planes forming a helical
configuration. The propagation vector of the magnetic spiral lies
along one of the three equivalent $<001>$ axes \cite{plumier:66}.
The transition to the AFM state at $T_\mathrm{N}$ is accompanied
by a structural transformation from cubic $Fd\bar{3}m$ to
tetragonal $I4_1/amd$ symmetry with a small contraction along the
$c$ axis of $c/a= 0.9999$ \cite{kleinberger:66}. Recent
neutron-diffraction and x-ray studies using synchrotron radiation
more precisely defined the low temperature phase identifying an
orthorhombic $Fddd$ symmetry \cite{hidaka:03}. Such a structural
transformation cannot originate from the ordinary Jahn-Teller
instability, because the orbital moment of the Cr$^{3+}$ ions is
quenched in a cubic crystal field. However, as is well documented
in literature, many chromium oxide and chalcogenide spinels
manifest structural instabilities accompanying the magnetic
ordering. In these compounds Cr$^{3+}$ reveals a half filled
$t_{2g}$ crystal-field ground state with almost zero spin-orbit
coupling. Although the oxide, sulfide, and selenide are dominated
by different exchange interactions, as indicated by their
Curie-Weiss (CW) temperatures, they reveal similar magnetic
transition temperatures into AFM states. In ZnCr$_2$O$_4$ with the
smallest Cr-Cr separation and a CW temperature of -390~K,  a
transition from a paramagnet with strong quantum fluctuations into
a planar antiferromagnet occurs at $T_\mathrm{N}$ = 12.5~K along
with a small tetragonal distortion \cite{lee:02,lee:00,hidaka:03}.
The oxide is governed by strong geometrical frustration of spins
coupled by direct AFM Cr-Cr exchange \cite{goodenough:60}. In
sulfide, ZnCr$_2$S$_4$, with a higher Cr-Cr separation, FM Cr-S-Cr
and AFM exchange interactions almost compensate each other
yielding a CW temperature close to zero \cite{hemberger:06}. Below
15 K, ZnCr$_2$S$_4$ undergoes a magnetic phase transition into an
incommensurate helical spin order similar to that of
ZnCr$_2$Se$_4$ \cite{hamedoun:86}. But at lower temperatures a
second commensurate collinear antiferromagnetic phase develops
which has a similar spin arrangement like the AFM ZnCr$_2$O$_4$.
At low temperatures both magnetic phases coexist
\cite{hamedoun:86}. The two subsequent antiferromagnetic
transitions in ZnCr$_2$S$_4$ at 15~K and 8~K are accompanied by
pronounced thermal and phonon anomalies. Due to strong spin-phonon
coupling both magnetic phase transitions induce a splitting of
phonon modes. The anomalies in the specific heat and thermal
expansion and phonon splitting observed at 8~K are strongly
suppressed by a magnetic field which supports the FM correlations.
This evidences the spin-driven origin of the structural
transformation related to strong competition of ferromagnetic and
antiferromagnetic exchange interactions \cite{hemberger:06}. In
ZnCr$_2$Se$_4$, with the highest Cr-Cr separation, the direct
exchange is almost suppressed and the spin arrangement follows
from the dominating FM $nn$ 90$^\circ$ Cr-Se-Cr exchange and the
additional AFM $nnn$ Cr-Se-Zn-Se-Cr and Cr-Se-Se-Cr exchange
interactions \cite{menyuk:66}. Therefore, one can expect a much
stronger influence of the magnetic field on the structural degree
of freedom. Here we use susceptibility, electron-spin resonance
(ESR), specific heat, and thermal expansion to probe the
spin-lattice correlations and to elucidate the origin of the
structural instability in the ZnCr$_2$Se$_4$ spinel.

Polycrystalline ZnCr$_2$Se$_4$ was prepared by solid-state
reaction from high purity elements at 1000~$^\circ$C. The single
crystals were grown by chemical transport reactions between 900
and 950~$^\circ$C from the polycrystalline material. X-ray
diffraction analysis of the powdered single crystals at room
temperature revealed a single-phase material with the cubic spinel
structure (see inset in Fig.~1a) with a lattice constant $a =
10.498(2)$~{\AA} and a selenium fractional coordinate $x =
0.260(1)$. The magnetic properties were studied using a commercial
SQUID magnetometer (Quantum Design MPMS-5) up to 50~kOe and a $dc$
extraction magnetometer (Oxford Instruments) up to 100~kOe. The
heat capacity was measured in a Quantum Design PPMS for
temperatures $2~\mathrm{K}~<~T~<~300~\mathrm{K}$ and in external
magnetic fields up to 70~kOe. The thermal expansion was measured
employing a capacitive method in fields up to 70~kOe. The ESR
studies were carried out with a Bruker ELEXSYS E500
CW-spectrometer at X-band frequency ($\nu$= 9.36~GHz) in a helium
gas flow cryostat (Oxford Instruments) for a temperature range
between 4.2 and 300~K. For the ESR experiments a thin disk with
the faces along the crystallographic (110) plane was cut from the
single crystal allowing for monitoring all three principal cubic
directions for the in-plane magnetic field.

\begin{figure}
\centering
 \includegraphics[width=0.40\textwidth,clip]{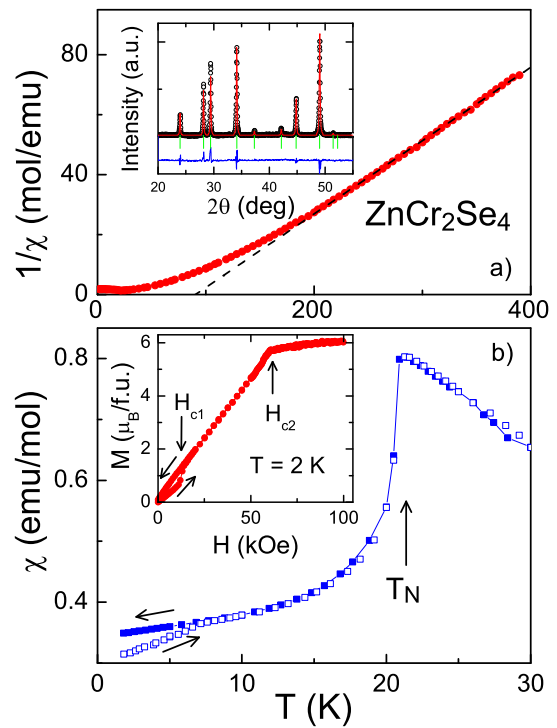}
 \caption{(color online) a) Inverse susceptibility vs temperature as measured
 in a ZnCr$_2$Se$_4$ single crystal at 10~kOe. The dashed line indicates
 a Curie-Weiss behavior. Inset: x-ray diffraction profile of the powdered single crystal.
 The measured intensities (open circles) are compared with the calculated profile
 using Rietveld refinement (solid line). Bragg positions of the normal
 cubic spinel structure are indicated by vertical bars and the
 difference pattern by the lower thin solid line; b) susceptibility
 vs temperature at low temperatures. The arrow indicates the magnetic
 phase transition at $T_\mathrm{N}$. Inset: Magnetization curve for a field applied
 along the $<001>$ direction at 2~K. The arrows indicates the critical
 fields $H_\mathrm{c1}$ and $H_\mathrm{c2}$ as described in the text.}
  \label{fig1}
\end{figure}

Figure ~\ref{fig1}a presents the inverse susceptibility
$\chi^{-1}$ vs temperature for a single crystalline sample. For
temperatures above 300~K, $\chi$ follows a Curie-Weiss law with a
positive CW temperature $\Theta =90$~K and an effective moment of
4.04~$\mu_\mathrm{B}$ close to the spin-only value of
3.86~$\mu_\mathrm{B}$ for Cr$^{3+}$ ions in a $3d^{3}$
configuration. Below 200~K,  $\chi^{-1}$(T) deviates from the CW
law indicating an increasing contribution of spin fluctuations. At
21~K the susceptibility shows a pronounced maximum with a steep
drop followed by a smooth continuous decrease towards lower $T$
(Fig.~1b). The temperature of 21~K marks the onset of long-range
antiferromagnetic order at $T_\mathrm{N}$ identified by neutron
diffraction studies \cite{plumier:66, hidaka:03}. At temperatures
below $T_\mathrm{N}$ the magnetization $M$ shows a change of slope
at a critical field $H_\mathrm{c1}$ of about 10~kOe characteristic
for a metamagnetic transition as shown in the inset in Fig.~1b for
$T$ = 2~K. This feature corresponds to the reorientation of
domains according to the nearly equivalent crystallographic
principal axes. The second critical field $H_\mathrm{c2}$ (which
for $T$ = 2~K is 65~kOe; see inset of Fig.1b) corresponds to the
breakdown of the helical spin arrangement. Beyond $H_\mathrm{c2}$
the magnetization $M$ reaches the full saturation of about 3
$\mu_\mathrm{B}$ per Cr ion.

\begin{figure}
\centering
 \includegraphics[width=0.40\textwidth,clip]{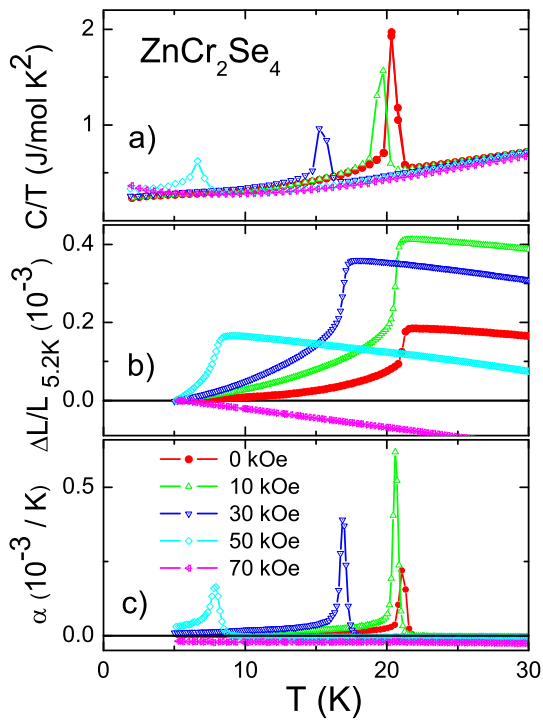}
 \caption{(color online) Temperature dependence of the heat capacity
 plotted as $C/T$ vs $T$ (a), of the thermal expansion $\Delta~L/L$
 (b) and of the thermal expansion coefficient $\alpha$ (c) for a single crystalline
 ZnCr$_2$Se$_4$ sample at different external magnetic fields between 0
 and 70~kOe applied along $<001>$ axis.}
  \label{fig2}
\end{figure}

Figure~\ref{fig2}a shows the specific heat in the representation
$C/T$ vs $T$ at different external magnetic fields. In the absence
of a magnetic field, the specific heat manifests a sharp anomaly
on approaching the N\'{e}el temperature $T_\mathrm{N}$.
Application of magnetic fields suppresses the peak in the specific
heat concomitantly shifting it to lower temperatures. The observed
peak in $C/T$ at $T_\mathrm{N}$, which is much sharper than for
conventional antiferromagnets, points towards a first-order
transition.  The anomaly in $C/T$ is fully suppressed by a
magnetic field of 70~kOe where the helical spin arrangement is
completely destroyed and the magnetization reaches the saturation
as discussed above. The entropy involved in the antiferromagnetic
transition, calculated by integrating
($C_\mathrm{0}$-$C_\mathrm{70~kOe}$)/$T$ over the transition
region, is 2.1~J mol$^{-1}$K$^{-1}$. This value is surprisingly
low, reaching only 9 \% of the full entropy of $2Rln4$ (23.05~J
mol$^{-1}$K$^{-1}$) expected for the complete ferromagnetic
alignment of the Cr spins. Assuming that the magnon contribution
to the specific heat for 0 and 70~kOe does not differ too much,
this anomalously low transition entropy reveals that the main part
of the total spin entropy is released already at temperatures much
higher than $T_\mathrm{N}$. This indicates strong spin
fluctuations in the paramagnetic regime characteristic for highly
frustrated magnets. Fig. 2a also documents how magnetic order can
be fully suppressed by an external magnetic field yielding a
quantum-critical point with a $T$ = 0 AFM phase transition. To our
knowledge this is one of the rare examples how quantum criticality
can be reached not only in metallic systems with competing
magnetic and Kondo-type interactions but also in insulating
geometrically frustrated magnets.

Figure 2b presents the thermal expansion $\Delta L/L$($T$) for a
single crystalline sample measured at different magnetic fields.
$\Delta L/L$ exhibits a strong drop at $T_\mathrm{N}$. Figure 2c
illustrates the respective variations of the thermal expansion
coefficient $\alpha=(1/L)dL/dT$. It manifests a steep narrow
maximum at $T_\mathrm{N}$ for zero-field which shifts to lower
temperature and broadens with increasing field similarly to the
anomaly in the specific heat. In contrast to the specific heat,
the amplitude of the maximum in $\alpha$ and the change of $\Delta
L/L$ at the transition temperature show a non-monotonous variation
with magnetic field. It increases with the field up to the
critical value $H_\mathrm{c1}$ of the metamagnetic transition and
than decreases for higher fields. Such a behavior for low fields
can be attributed to the reorientation of magnetic domains with
three different nearly equivalent $<001>$ axes being the spiral
propagation direction in agreement with the magnetization
measurements \cite{nogues:85}. The thermal expansion is strongly
anisotropic being nearly three times larger along the $<001>$ axis
compared to the $<111>$ axis.

\begin{figure}
\centering
 \includegraphics[width=0.40\textwidth,clip]{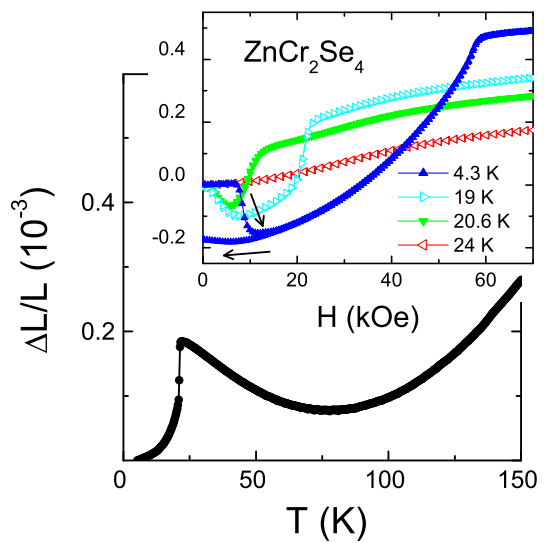}
 \caption{(color online) Thermal expansion $\Delta L/L$($T$) of a
 ZnCr$_2$Se$_4$ single crystal for an extended temperature range
 illustrating the negative thermal expansion between $T_\mathrm{N}$ and 75 K.
 Inset: $\Delta L/L$($H$) at several temperatures below and above $T_\mathrm{N}$.}
 \label{fig3}
\end{figure}

The other important feature of $\Delta L/L$ is a strong negative
thermal expansion observed below 75~K down to $T_\mathrm{N}$ as
demonstrated in Figure~\ref{fig3}. It is interesting to note that
in an external field of 70 kOe a strictly constant negative
thermal expansion evolves at the lowest temperature of our
measurements (5.2~K), resulting in an approximately linear
decrease of the cell dimension with increasing temperature up to
75~K (see Figs.~\ref{fig2}b and ~\ref{fig2}c). Negative thermal
expansion could result from the geometrical frustration of the
lattice degree of freedom \cite{ramirez:03} and is usually
explained by highly anharmonic vibrational modes \cite{ernst:98},
which in the case of ZnCr$_2$Se$_4$ have to result from strong
coupling of the phonons to the spin degree of freedom. The
magnetostriction, i.e. the field dependence of the lattice
expansion, below and above $T_\mathrm{N}$ is illustrated in the
inset of Fig.~\ref{fig3}. The monotonous increase of $\Delta
L/L$$(H)$ observed above $T_\mathrm{N}$ changes into an initial
decrease up to the metamagnetic transition at 10~kOe followed by a
stronger increase up to the saturation field at temperatures below
$T_\mathrm{N}$. Note the unusually high value of the
magnetostriction which is comparable to that observed in giant
magnetostrictive materials with strong spin-orbital coupling
\cite{mahendiran:03}. However, it is necessary to remind that a
strong spin-orbital coupling is not expected here, as we
additionally proved by the electron-spin resonance measurements
discussed below.

\begin{figure}
\centering
 \includegraphics[width=0.38\textwidth,clip]{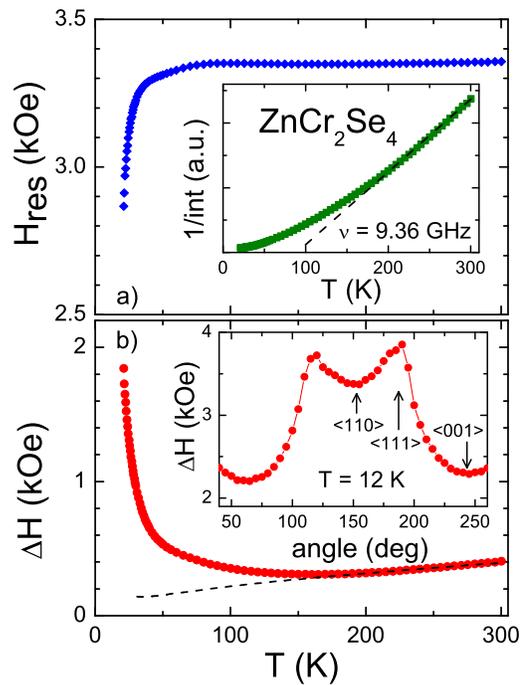}
 \caption{(color online) Temperature dependence of the resonance field
 (a) and linewidh (b) for a single crystalline ZnCr$_2$Se$_4$
 disc with (110) plane orientation. The dashed line indicates
 the expected behavior for an exchange coupled system.
 Upper inset: inverse intensity of the ESR line vs $T$ showing a Curie-Weiss
 dependence at high temperatures. Lower inset: angular dependence of the linewidth
 at 12~K revealing magnetocrystalline anisotropy.}
 \label{fig4}
\end{figure}

To gain insights into the local magnetic properties of
ZnCr$_2$Se$_4$ we measured the temperature dependence of the ESR
absorption of the Cr spins which consists of a single
exchange-narrowed Lorentz line in the paramagnetic regime.
Figure~\ref{fig4} shows the temperature dependence of the
resonance field and linewidth for the single crystalline sample in
form of a disc with (110) plane orientation for temperatures
$21~\mathrm{K}~<~T~<300~\mathrm{K}$. From the resonance field we
derived an asymptotic $g$-value of 1.996 at high temperatures
close to the spin-only value in good agreement with the results of
the paramagnetic susceptibility. This additionally indicates only
minor spin-orbit coupling typical for Cr$^{3+}$ ions
\cite{abragam:86}. For lower temperatures and approaching the
magnetic phase transition the resonance field strongly decreases
as often observed close to the onset of AFM order due to the
opening of the excitation gap. The $T$-dependence of the intensity
at resonance absorption compares well with the bulk susceptibility
indicating the same Curie-Weiss law at high temperatures (inset in
Fig.~4a). The linewidth, which is a measure of the spin
correlations, strongly increases when approaching the AFM ordering
temperature. For an exchange coupled spin system outside the
critical regime and above the phase transition, the temperature
dependence of the linewidth should be proportional to the $\Delta
H_\infty$/($T\chi$) where $\Delta H_\infty$ denotes the asymptotic
high-temperature linewidth \cite{huber:99}. Deviations from this
expected behavior indicated by the dashed line in Fig.~\ref{fig4}b
set in already below 200 K signaling significant spin fluctuations
due to strong exchange interactions despite the low ordering
temperature. This underlines the competition of the FM and AFM
interactions evident from the corresponding high-temperature
deviation of the susceptibility from the CW law. Note that no
anisotropy of the spectra is visible in the paramagnetic regime.
Below $T_\mathrm{N}$ a broad absorption band is observed centered
on zero magnetic field. The angular dependence of the
corresponding linewidth is shown in the inset of Fig.\ref{fig4}b.
It reflects a cubic anisotropy with a maximum at $<111>$ and
minima at $<110>$ and $<001>$ axes indicating strong coupling of
the magnetization to the lattice in agreement with the earlier
higher frequency ESR results \cite{siratori:71}.

The obtained data reveal a strong spin-lattice coupling and
provide experimental evidence for a spin-driven origin of the
structural transformation from the high-temperature cubic phase to
a lower symmetry below the magnetic transition at $T_\mathrm{N}$.
The dominant magnetic coupling mechanism in ZnCr$_2$Se$_4$ is
superexchange which includes $nn$ FM Cr-Se-Cr and $nnn$ AFM
Cr-Se-Zn-Se-Cr or Cr-Se-Se-Cr exchange interactions
\cite{menyuk:66}. These competing exchange interactions establish
the complex incommensurate spin order in this compound. The FM
90$^\circ$ Cr-Se-Cr exchange governs the ferromagnetic order in
the (001) planes. The AFM exchange is probably responsible for the
spin arrangement between the adjacent (001) planes. An external
magnetic field changes the balance between FM and AFM interactions
enhancing the ferromagnetic correlations and thus reduces the
angle between the spins in the adjacent FM planes. In agreement
with our observation of the concomitant reduction of the anomalies
of specific heat and thermal expansion by an external magnetic
field this corroborates the interpretation of the low-temperature
structural symmetry breaking in ZnCr$_2$Se$_4$ as due to bond
frustration caused by competing exchange. This results in an
extremely large influence of the magnetic field on the structural
transition. The negative thermal expansion indicates a high degree
of frustration of this highly symmetric lattice in the
paramagnetic regime. This frustrated lattice is strongly receptive
to weak perturbations which are primarily induced by magnetic
order at $T_\mathrm{N}$. The coupling between the magnetism and
lattice may be realized via an exchange-striction mechanism
similar to a spin-Peierls transition. However, compared to the 3D
spin Jahn-Teller transition in chromium oxide spinels
\cite{lee:00,tchernyshyov:02,yamashita:00} and to the 1D
spin-Peierls transition in CuGeO$_3$ \cite{hase:93} we observed a
much stronger effect of an external magnetic field on the
structural transition which we attribute to the strong bond
frustration.

In conclusion, we investigated the bond frustrated AFM spinel
ZnCr$_2$Se$_4$ and found pronounced anomalies in the specific heat
and thermal expansion at the onset of the antiferromagnetic
helical order at the N\'{e}el temperature. Our results reveal
strong spin-phonon coupling that generates the low-temperature
structural instability in this compound. The observed negative
thermal expansion is suggested to result from the high frustration
of the lattice degrees of freedom.  A colossal magnetostriction
comparable to giant magnetostrictive materials is found despite
the absence of strong spin-orbital coupling of the half-filled
Cr$^{3+}$ $t_{2g}$ electronic state. An extremely strong
suppression of the anomalies in the specific heat and thermal
expansion by magnetic fields suggests a spin-driven origin of the
structural transformation and proximity to a quantum critical
point.

We are grateful to Dana Vieweg and Thomas Wiedenmann for
experimental assistance. This work was supported by BMBF via
VDI/EKM, FKZ 13N6917-B and by DFG within SFB 484 (Augsburg).

\end{document}